\documentclass[12pt]{article}
\usepackage{group21}
\title{Global restrictions to the Mixing Angle $\tw$}
\author{\underline{M. Calixto}\adr{1}, \ V. Aldaya\adr{1,2} \and\ 
J. Guerrero\adr{1,3}}
\address[1]{Instituto ``Carlos I" de F\'\i sica Te\'orica 
y Computacional, Facultad de Ciencias, Universidad de Granada,   
Campus de Fuentenueva, Granada 18002, Spain. E-mail: calixto@goliat.ugr.es; 
\,\,http://www.ugr.es/\~{}valdaya/mphys}

\address[2]{IFIC, Centro Mixto Universidad de Valencia-CSIC, Burjasot
              46100-Valencia, Spain. E-mail: valdaya@goliat.ugr.es}

\address[3]{Dipartimento de Scienze Fisiche, Mostra d$'$ Oltremare, Pad. 19, 
80125 Napoli, Italy. E-mail: guerrero@goliat.ugr.es}

\def\ni{\noindent}

\def\be{\begin{equation}}
\def\ee{\end{equation}}
\def\bea{\begin{eqnarray}}
\def\eea{\end{eqnarray}}
\def\ba{\begin{array}}
\def\ea{\end{array}}
\def\tw{\theta_W}
\def\gg{SU(2)\otimes U(1)_{Y}}

\begin{document}
\maketitle

\section{Introduction}

In spite of its firm hold  in particle physics, 
the gauge group $\gg$ proves not to be the most 
appropriate to describe the unification of weak and 
electromagnetic interactions. In fact, a look at the structure of $\gg$ tells us that both, the electric 
charge generator $Q$ and its corresponding electromagnetic 
gauge field $A_\mu(x)$, are not basic constituents of this gauge group. 
Rather, the ``weak isospin'' $T_1,T_2,T_3$  (with  its 
corresponding gauge fields  $A^{(1)}_\mu,A^{(2)}_\mu,A^{(3)}_\mu$) 
and the hyphercharge $T_4\equiv Y$ 
(which provides another gauge field  $A^{(4)}_\mu$) are its basic generators.
Nevertheless, there is a possibility of defining a proper electric 
charge through the Gell'Mann-Nishijima relationship 
\begin{equation}
Q\equiv {\tilde{T}}_4=T_3+\frac{1}{2}T_4\label{G-N}
\end{equation}
\noindent and the associated electromagnetic field through the mixture
\begin{equation}
A_\mu\equiv \tilde{A}^{(4)}_\mu= R^4_3A^{(3)}_\mu+R^4_4A^{(4)}_\mu\,.\label{electro}
\end{equation}
\ni It can be seen that {\it canonical independence} requirements 
in the Poisson brackets of the gauge field theory 
(see e.g. \cite{Bernstein}) force us to define  the counterpart 
mixture, that which leads to the neutral vectorial boson
\be
Z^{(0)}_\mu\equiv\tilde{A}^{(3)}_\mu= R^3_3A^{(3)}_\mu+R^3_4A^{(4)}_\mu\,,\label{boson}
\ee
\ni and force the transformation in (\ref{electro}) and (\ref{boson}) to be ortogonal, i.e.,
\be
\begin{array}{c}\hbox{canonical}\\ \hbox{independence}\\\hbox{conditions}\end{array} \Rightarrow \left(\begin{array}{cc}R^3_3&R^3_4\\R^4_3&R^4_4\end{array}\right)=
\left(\ba{cc} \cos\tw &-\sin\tw \\ \sin\tw &\cos\tw \ea\right)\,.\label{rot}
\ee 
\ni Thus, the neutral weak charge 
would be given by a mixture  
\begin{equation}
T_0\equiv {\tilde{T}}_3=C^3_3T_3+C^4_3 T_4\,,\label{G-Nc}
\end{equation}
\ni where the coefficients $C^3_3, C^4_3$ depend on the 
mixing angle $\tw$. To be more precise, the invariance 
of the connection 1-form $\Lambda=(g)^a_bT_aA^{(b)}_\mu dx^\mu$ under 
a general Lie algebra  transformation 
 $\tilde{T}_a=C^b_aT_b$, induces a 
transformation in the gauge fields of  the form
\be
\tilde{A}^{(a)}_\mu=(\tilde{g}^{-1})^a_b(C^{-1})^b_c(g)^c_dA^{(d)}_\mu\equiv 
R^a_d A^{(d)}_\mu\label{gt}
\ee
\ni where $(g)=\hbox{diag}(r,r,r,r')$ and $(\tilde{g})=\hbox{diag}(\tilde{r}
,\tilde{r},\tilde{r},\tilde{r}')$ are the initial and final (bare) coupling-constant matrices. For the specific Lie algebra transformation given 
in (\ref{G-N},\ref{G-Nc}), the identification of  $R^a_d$ in (\ref{gt}) with 
the rotation  (\ref{rot}) gives the value of 
$\sin^2\tw=\frac{\tilde{r}'{}^2}{4r^2}(\equiv\frac{e^2}{g^2}$ in 
the standard notation) in terms of the quotient 
of final and initial coupling constants. This characterization of 
the mixing angle differs from the more conventional  expression of 
$\sin^2\tw=1-M_W^2/M_Z^2$ in 
terms of the masses of the vectorial bosons $Z^{(0)}_\mu$ and 
$W^\pm_\mu\equiv\frac{1}{\sqrt{2}}(A^{(1)}_\mu\pm iA^{(2)}_\mu)$. This other 
characterization of $\tw$ is  strongly related to the Symmetry 
Breaking mechanism and 
arises as an orthogonal transformation which diagonalizes 
a symmetric mass matrix. Even though these two definitions 
of $\tw$ are usually identified, there are conceptual 
differences, as  in fact pointed out in \cite{Veltman}. 
We shall go further and show that, whereas the  quantity 
$1-M_W^2/M_Z^2$ 
is a free parameter which has to be fixed by experiments,  
there are strong mathematical restrictions to the value of 
${e^2}/{g^2}$ so as to fix it to 
${1}/{2}$ ($\tw=\pi/4$).
\section{Global restrictions to $\tw$}

If we disregard  the global structure of the gauge group,  
then any one-to-one Lie algebra transformation 
 $\tilde{T}_a=C^b_aT_b$ (as that given in (\ref{G-N},\ref{G-Nc}))  
would clearly be allowed for arbitrary coefficients $C^b_a$. However, the increasing 
importance  in gauge theories of some basic topological issues (for example, the existence or not of monopoles and solitons, 
topological properties of the Symmetry Breaking, the 
Bohm-Aharonov effect itself, etc.) demands a 
revision of some local (Lie algebra level) transformations. 
In fact,  since the group $\gg$ is non-simply connected, there are 
strong restrictions to the 
number of globally exponentiable Lie algebra transformations 
\cite{Chevalley};  in other words, not all the Lie algebra 
isomorphisms can be realized as 
the derivative of  global homomorphisms of the corresponding gauge group. 
Without going into details (see \cite{articulo} for more information), 
we can say that 
 the mere embedding of the electromagnetic subgroup 
$U(1)_Q$ in the torus $T^2=U(1)_{T_3}\otimes U(1)_Y$, as suggested by 
(\ref{G-N},\ref{G-Nc}), imposes non-trivial restrictions 
(rational values) to the 
tangent of the integral curves ({\it closed} geodesics) associated with its generator. That is, 
the coefficients $C^3_3,C^3_4$ in (\ref{G-Nc}) have to be rational 
numbers, this fact 
leading to fractional values for $\tan^2\tw={n}/{m}$, 
according to (\ref{gt}). Then, the 
additional requirement that $U(1)_{T_3}$ be a subgroup of $SU(2)$  imposes 
further, severe restrictions so as to fix $\tan^2\tw=1$. The corresponding 
group homomorphism proves to be 
\be
\gg \buildrel \tw=\pi/4\over\longrightarrow 
\left(\gg\right)/{Z_2}\simeq U(2)\,.\label{homo}
\ee
\section{Conclusions}
In summary, the only global homomorphism  
from $\gg$ to a 
locally isomorphic group defining a proper rotation on the gauge fields 
compatible with the Gell'Mann-Nishijima relationship (thus providing an electric 
charge) is the homomorphism (\ref{homo}), which leads 
to the value $\sin^2\tw =1/2$ for the mixing angle. 
This means that there is only one final coupling constant --essentially the electric charge, i.e., 
$e\equiv \tilde{r}' = \sqrt{2}r(\equiv g/\sqrt{2}$ in the standard notation)--,
even though the gauge group ($U(2)$) is not a simple group. According to 
general settings, however, the theory must contain a coupling constant 
for each simple
or abelian term in the Lie algebra decomposition of the gauge group. An immediate 
conclusion is that the {\it assignment of  coupling constants} should 
be done according to factors in the direct product
decomposition of the {\it group}, rather than the algebra. 

With regard to the structure of the currents, 
this particular value of $\tw$  means that the neutral weak 
current is pure V$-$A for the 
neutrino and pure V+A for the lepton.

In the light of the difference between the value of ${e^2}/{g^2}={1}/{2}$ 
(previous to any mechanism intended to supply masses) and the 
experimental value of $1-{M^2_W}/{M^2_Z}\approx 0.23$, only the hope 
remains that our characterization of $\tw$ really corresponds to an 
{\it asymptotic
limit} (high energies), or to that 
state of the Universe in which the electroweak interaction was not yet 
``spontaneously broken", i.e. the masses of the vector bosons are zero and 
therefore the quantity $1-{M_W^2}/{M_Z^2}$ makes no physical sense. 
In any case,
our results provide strong support to the idea that those two  quantities  
cannot be directly identified, as it was in fact  
pointed out in \cite{Veltman}.
\section*{Acknowledgments}
This work was partially supported by the DGICYT. M. Calixto thanks the Spanish MEC for a FPI grant.


\begin{thebibliography}{99}
\bibitem{Bernstein}J.~ Bernstein,  Rev. Mod. Phys. {\bf 46}, 7 (1974). 
\bibitem{Veltman} G.~ Passarino and M. ~Veltman,  Phys. Lett. {\bf B237}, 537 (1990).
\bibitem{articulo}V. ~Aldaya, M.~ Calixto and J.~ Guerrero, 
Int. J. of Theor. Phys, {\bf 35}, 1901 (1996).
\bibitem{Chevalley} C.~ Chevalley, {\it Theory of Lie Groups}, Princeton 
University Press (1946).
\end{thebibliography}
\end{document}